\documentclass[conference]{IEEEtran}
\usepackage{amsfonts,amsmath,graphicx}
\usepackage{float}
\usepackage{color}
\usepackage{amsthm}
 
\newtheorem*{remark}{Remark}
 
\def\x{{\mathbf x}}
\def\N{{\mathbb N}}
\def\R{{\mathbb R}}

\def\cM{{\cal M}}
\def\eye{{\mathbf I}}
\def\ba{{\mathbf a}}
\def\cN{{\cal N}}

\def\ind{{\mathbf 1}}

\title{Bayesian autoregressive spectral estimation}

\author{
\IEEEauthorblockN{Alejandro Cuevas$^1$\qquad Sebastián López$^1$ \qquad Danilo Mandic$^{2}$\qquad Felipe Tobar$^{3,4}$}
\vspace{1em}
\IEEEauthorblockA{
\textit{$^1$Department of Mathematical Engineering, Universidad de Chile}\\
\textit{$^2$Department of Electrical and Electronic Engineering, Imperial College London}\\
\textit{$^3$Center for Mathematical Modeling, Universidad de Chile}\\
\textit{$^4$Initiative for Data \& Artificial Intelligence, Universidad de Chile}
}

}

\begin{document}
\maketitle
\begin{abstract}
Autoregressive (AR) time series models are widely used in parametric spectral estimation (SE), where the power spectral density (PSD) of the time series is approximated by that of the \emph{best-fit} AR model, which is available in closed form. Since AR parameters are usually found via maximum-likelihood, least squares or the method of moments, AR-based SE fails to account for the uncertainty of the approximate PSD, and thus only yields point estimates. We propose to handle the uncertainty related to the AR approximation by finding the full posterior distribution of the AR parameters to then propagate this uncertainty to the PSD approximation by \emph{integrating out the AR parameters}; we implement this concept by assuming two different priors over the model noise. Through practical experiments, we show that the proposed Bayesian autoregressive spectral estimation (BASE) provides point estimates that follow closely those of standard autoregressive spectral estimation (ASE), while also providing error bars. BASE is validated against ASE and the Periodogram on both synthetic and real-world signals.

\end{abstract}
\begin{IEEEkeywords} Spectral estimation, autoregressive models, Bayesian inference, Markov chain Monte Carlo
\end{IEEEkeywords}%

% Add introduction
%!TEX root = Bayesian_AR_LACCI2021.tex

\section{Introduction}
\label{sec:intro}

In Signal Processing, spectral estimation (SE) refers to the determination of the energy that is contributed by each component in a time series. This work focuses, in particular, on the spectral representation given by the Fourier power spectral density (PSD). In practice, the challenge is to estimate the PSD from a set of noisy observations of the time series; this is seen in applications such as biomedical engineering \cite{mandic_eeg} and telecommunications \cite{5496152}. Classical methods for computing the PSD can be roughly divided in two categories: parametric and nonparametric. Parametric methods impose a generative model on the data such as autoregressive moving average (ARMA) models \cite{book:arma_psd}, sums of sinusoids (Lomb-Scargle \cite{lomb}), Bernstein polynomials \cite{choudhuri_2004} or Gaussian mixtures \cite{tobar17}. On the other hand, nonparametric methods for spectral estimation do not assume any structure on the data and are thus more flexible, yet they do not discriminate the signal from the noise, which critically affects the spectral estimates. Typical examples of the nonparametric SE methods are the classic Periodogram \cite{Williams:1997:DSP:550591} which can be considered as a \emph{histogram of frequencies} as well as recent Bayesian nonparametric approaches \cite{tobar15, tobar18,Wang_2012}.

Even though the randomness conveyed by time series observations is evident, dealing with the uncertainty related to the data-driven computation  of the PSD is not a standard practice within SE. In parametric approaches, the model fit to the data through an optimisation procedure, e.g. using least squares or maximum likelihood. Therefore, the PSD is approximated deterministically, by a point estimate, and all the information about the mismatch between the chosen model and the data, which could lead to probabilistic PSD estimates, is unfortunately usually neglected. 

To address this void in the SE literature, which is particularly detrimental in noisy data and model mismatch, we aim to cater for uncertainty in PSD estimates finding a \emph{distribution over PSDs}, rather than point estimates. This can be achieved by equipping the existing framework for parametric spectral estimation with a Bayesian treatment. Instead of committing to a point-estimate of parameters of the time-series model, we can find their posterior distribution, thus constructing a posterior distribution over models for time series. Then, we can \emph{transport} this distribution via the Fourier transform to find the posterior distribution over PSDs, where \emph{posterior} in this case is in the sense of conditional to the time series observations. This rationale becomes of particular importance for model misspecification, since, strictly speaking, a parametric model may always be considered to be misspecified for real data and therefore the distribution of models allows us to account for the limited capacity of the parametric representation. 

In this work, we apply this generic idea to the particular class of autoregressive time series models, thus our proposed method is referred to as Bayesian autoregressive spectral estimation (BASE). In particular, we propose two models, termed Model I and Model II which difference stems from the noise prior considered: Model I uses a half-Normal prior while Model II uses an Inverse-Gamma (and thus conjugate) prior. Both models assume a normal prior over the AR weights and Model I is implemented using  Markov chain Monte Carlo (MCMC) \cite{Andrieu2003} due to the nonconjugate prior. Via numerical simulations, we validate the proposed BASE method and the two models proposed on two synthetic signals and a real world example, these experiments show the advantage of the proposed BASE in different settings against its deterministic counterparts.

% Add methodology part
%!TEX root = Bayesian_AR_LACCI2021.tex

\section{Background: Parametric spectral estimation using an AR model}

A discrete-time sequence $\{x_t\}_{t\in\N}$ is called a $p$-order autoregressive process, denoted AR($p$), if it is given by the relationship \cite{Williams:1997:DSP:550591}
\begin{equation}{\label{eq:arma}}
 x_t = \sum_{k=1}^{p}a_k x_{t-k}  + \epsilon_t,
\end{equation}
where $p\in\N$, $[a_1,\ldots, a_p]\in\R^p$ are the (autoregressive) parameters and $\{\epsilon_t\}_{t\in\N}$ is a noise process. We consider the noise to be Gaussian independent and identically distributed random variables with zero mean and variance $\sigma^2$. 

\subsection{Power spectral density of AR($p$) processes}
\label{ssec:ar_psd}
AR($p$) processes are ubiquitous in parametric spectral estimation as they have closed-form Fourier PSDs \cite{Box:1990:TSA:574978} given explicitly by the AR parameters. Specifically, using the $Z$-transform \cite{book:arma_psd} the PSD of AR($p$) can be expressed by
\begin{equation}
 S_x(\xi) =  \frac{\sigma^2}{|1 - \sum_{k=1}^{p} a_k e^{-i2\pi \xi k}|^2}.
 \label{eq:arma_psd}
\end{equation}

The modes (or peaks) of the PSD in eq.~\eqref{eq:arma_psd}, i.e., the frequencies where the AR process convey more power, are given by the roots of the denominator. These roots are the poles of the dynamical system that generates the AR($p$) process $\{x_t\}_{t\in\N}$. Since we can regard the model order $p$ as the degrees of freedom of the parametric PSD for the AR model, the larger the $p$ the more flexible the PSD. This is in line with the interpretation in the temporal domain, where a model with more lags can cater for more complex time series and thus represents a more general model.

The fact that the PSD above is directly parametrised by $\{a_1, \ldots, a_p, \sigma_n \}$ is perhaps the main motivation for using the AR model for spectral estimation. However, even though there have been Bayesian approaches to fit the AR parameters they have not been applied to the SE problem. In practice, the literature shows that the use of AR parameters in SE has been mainly deterministic, e.g., by computing the PSD in eq.~\eqref{eq:arma_psd} with a plug-in approximation of the AR parameters such as those found by least squares, the method of moments, or maximum likelihood. Such an approach, does not allow for modelling the uncertainty related to the peaks of the resulting PSD, due to noisy data, or to account for model misspecification (i.e., incorrect model order) which can lead to either peak splitting (over-estimation) or onto fewer peaks (under-estimation).

% Add model part
%!TEX root = Bayesian_AR_LACCI2021.tex

\section{Bayesian spectral estimation of autoregressive signals}

Our aim is to compute the posterior distribution of the PSD for the AR case, thus providing a  natural account for model uncertainty, which arises from i) the consideration of a finite dataset, ii) model mismatch, iii) noisy observations. From now on, we will refer to the standard (deterministic) autoregressive spectral estimation approach as ASE and to the proposed Bayesian counterpart as BASE.

\subsection{Generic Bayesian parametric spectral estimation}

The posterior distribution of $S_x$,  the PSD of a stationary stochastic process $\{x\}_{t\in\N}$, conditional to a set of observations $\x=[x_1,x_2,\ldots,x_n]$, can be computed by integrating out the model of the time series. Denoting the space of all possible time series models by $\cM$, the posterior PSD is given by 
\begin{align}
	p(S_x|\x) & = \int_\cM p(S_x,M|\x)dM\label{eq:post1}\\
	          & = \int_\cM p(S_x|M)p(M|\x)dM,\nonumber
\end{align}
where the last expression only takes the (reasonable) assumption that the PSD, $S_x$, and the observations, $\x$, are conditionally independent given the model $M$. This follows from the fact that  the data cannot say more about the PSD than the model itself.

Notice that the above integral is defined over the entire, possibly infinite-dimensional, model space $\cM$. To calculate this integral we need to assume a structure for the time-series models to be considered; this can be achieved by choosing a finite-dimensional prior over the models, namely $p(M)$. A straightforward way to choose this prior is to first assume a parametrisation over models, say $M=M_\theta,\ \theta\in\R^d$, and then choose a prior over the parameters $p(\theta)$. This construction is know as the push-forward measure \cite{halmos1976measure} that transports a distribution over the finite-parameter, $\theta\in\R^d$, towards the space of models through the mapping $\theta \mapsto M_\theta$.

With this parametrisation, the posterior in eq.~\eqref{eq:post1} can be expressed with respect to the model parameters, $\theta$, via the change of variable theorem as 
\begin{align}
	p(S_x|\x) & = \int_\Theta p(S_x|\theta)p(\theta|\x)d\theta,\label{eq:post2}
\end{align}
where the integration is now performed over the (finite) parameter space, that is, $\Theta = \R^d$.

Under the assumption that the process $\{x_t\}_{t\in\N}$ is stationary and given by a model $M_\theta$, its PSD is uniquely defined by the model's parameters $\theta$ and thus the distribution $p(S_x|\theta)$ is a Dirac measure supported on a single points on the space of PSDs. Therefore, sampling from the posterior over PSDs, $p(S_x|\x)$, is straightforward: simply sample a parameter from the posterior distribution over parameters $\theta^*\sim p(\theta|\x)$, and then map this parameter sample to PSDs according to 
\begin{equation}
	\theta^* \rightarrow M_{\theta^*} \rightarrow S_x=\text{PSD}(M_{\theta^*}),
	\label{eq:PSDmap}
\end{equation}
where $\text{PSD}(M)$ denotes the PSD corresponding to the model $M$. 

In order to perform this procedure, we need to choose: i) a model space $M_\theta$, ii) a prior over the parameters $p(\theta)$, and iii) a sampling procedure. We refer to these in the remaining part of this section.

\subsection{Model I: Half-normal noise} 
\label{sec:model_one}

We shall choose the model space as that of AR models owing to their appealing properties for spectral estimation outlined in Section \ref{ssec:ar_psd}. Furthermore, we will assume that all parameters in the AR model are independent, that $a_{1:p}$,  $p\in\N$, are Normally distributed and the noise variance $\sigma^2$ follows a half-Normal prior. Thus, the priors are
\begin{align}
	p(a_{1:p}) = \cM\cN(0,\sigma_a^2 \eye_p)\\
	p(\sigma^2) = \text{half-}\cN(0,\sigma_\epsilon^2 ),
\end{align}
where the half-Normal distribution corresponds to the multiplication between a Normal distribution and an indicator function $\ind_{\{\sigma^2\geq0\}}$, and then properly renormalised (i.e., multiplied by 2). 

The choice of normality for the autoregressive parameters follows from the fact that the Gaussian prior is conjugate to the (conditional) likelihood as verified in the next section, and thus the posterior is also normally distributed. 

Regarding the definition of the likelihood, given a set of observations $\{x_{1:T}\}$, $T\in\N$ consider the \textbf{conditional} likelihood of the AR($p$), that is, the expression $p(x_{p+1:T}|x_{1:p},\theta)$, where $\theta = [a_1,\ldots,a_p,\sigma^2]^\top$ denotes all model parameters and we consider $\{x_{1:p}\}$ to be fixed and not part of the generative model for simplicity of presentation. 

Following eq.~\eqref{eq:arma}, the conditional likelihood of the AR($p$) model is given by
\begin{align}
	p(x_{p+1:T}|x_{1:p},\theta) 
	&= \prod_{\tau =p+1 }^T p(x_\tau|x_{\tau-1:\tau-p},\theta) \\
	&= \prod_{\tau =p+1 }^T \frac{1}{\sqrt{2\pi\sigma^2}} \exp\left(\frac{-(x_\tau-\ba^\top \x_{\tau-1})^2}{2\sigma^2}\right),\nonumber 
\end{align}
where we have adopted the notation $\ba = [a_1,\ldots,a_p]$ and $\x_{\tau-1} = [x_{\tau-1},\ldots,x_{\tau-p}]$. 

The conditional likelihood is thus Gaussian on the AR parameter vector $\ba$, meaning that the Gaussian prior on $\ba$ established on the previous section is conjugate to the conditional likelihood and results on the following Gaussian posterior 
\begin{align}
	p(\theta|x_{1:p},x_{p+1:T}) & = \frac{p(x_{p+1:T}|x_{1:p},\theta) p(\theta) }{p(x_{p+1:T},x_{1:p})}
	\label{eq:posterior_dist}\\
	&\propto p(x_{p+1:T}|x_{1:p},\theta) p(\ba)p(\sigma^2)\nonumber\\
	&\propto \prod_{\tau =p+1 }^T \frac{1}{\sigma} \exp\left(\frac{-(x_\tau-\ba^\top \x_{\tau-1})^2}{2\sigma^2}\right)\nonumber\\
	&\quad\cdot \exp\left(\frac{-1}{2\sigma_a^2}a^Ta\right) \exp\left(\frac{-\sigma^2}{2\sigma_\epsilon^2}\right).\nonumber
	\end{align}

\begin{remark}
Since the complete (rather than conditional) likelihood is given by 
\begin{equation}
	\label{eq:likelihood}
	p(x_{1:T}|,\theta) =  p(x_{p+1:T}|x_{1:p},\theta) p(x_{1:p}|\theta),
\end{equation}
it requires calculating the distribution $p(x_{1:p}|\theta)$; this is challenging since it involves the infinite-order moving-average representation of the AR($p$) process. Therefore, under the assumption that the first $p$ observations are fixed they become independent from $\theta$, and the density $p(x_{1:p}|\theta)$ becomes a constant for theta in eq.~\eqref{eq:likelihood}. A similar reasoning applies for the marginal distribution $p(x_{p+1:T},x_{1:p})$. Notice that the discrepancy between the conditional and complete likelihoods becomes negligible for increasing amounts of data (i.e., $T\gg 1$). For this reason, we will consider only the conditional likelihood. 
\end{remark}

With this choice, generating a sample, $\theta^*$, from the posterior requires the evaluation of $p(x_{p+1:T}|x_{1:p},\theta) p(\ba)p(\sigma^2)$. Then, this sample can be used to compute the PSD explicitly, according to eq.~\eqref{eq:arma_psd}. This procedure avoids explicit computation of the model $M_\theta$, since the PSD is computed directly from the parameters.

Notice that, by construction, all the posterior PSD samples will have the form as in eq.~\eqref{eq:arma_psd} which means that no probability will be assigned to expressions that do not follow this form. This advantage of the parametric approach to SE results in that, using a set of samples of PSDs constructed as explained above, we can numerically compute the mean PSD and error bars conditional to a set of signal values $\x$.

\subsection{Model II: Inverse-Gamma noise} 
\label{sec:model_two}

In order to derive a closed-form expression for the posterior over the AR parameters, we can assume the noise variance $\sigma^2$ follows an Inverse-Gamma prior distribution, rather than the Half-Normal one in the previous section. This posterior is known as the Normal-Inverse-Gamma distribution. 

Specifically, let us consider the following priors
\begin{align}
	p(a_{1:p}|\sigma^2) &= \cM\cN\left(\mu_0,\dfrac{\sigma^2}{\lambda} \eye_p\right)\\
	p(\sigma^2) &= \Gamma^{-1}(\alpha, \beta),
\end{align}
where $\eye_p$ denotes the $p-$dimensional identity matrix.
Then the posterior distribution for $\theta=(a_{1:p},\sigma^2)$ is proportional to a Normal-Inverse-Gamma:
\begin{equation}
    p(\theta|x_{1:T}) \propto \cN\Gamma^{-1}(\mu, \Sigma^{-1},\overline{\alpha},\overline{\beta}),
\end{equation}
where, using the notation $X_{p+1:T} = [x_{i-1:i-p}^\top]_{i=p+1}^{T}$,
\begin{align*}
    \mu &= \Sigma(\lambda\mu_0 + X_{p+1:T}^\top x_{p+1:T})\\
    \Sigma^{-1} &= \lambda\eye_p + X_{p+1:T}^\top X_{p+1:T}\\
	\overline{\alpha} &= \alpha + \dfrac{T-p}{2}\\
    \overline{\beta} &= \beta + \dfrac{\lambda\mu_0^\top\mu_0 + x_{p+1:T}^\top x_{p+1:T} - \mu^\top\Sigma^{-1}\mu}{2}.
\end{align*}

From this Norma-Inverse-Gamma posterior, we can sample the parameters $(a_{1:p},\sigma^2)$ simply by specifying the hyper-parameters $(\mu_0, \lambda, \alpha, \beta)$ and using standard sampling techniques. Our conjecture is that avoiding the MCMC stage required above can result in faster posterior computations.

The MAP estimators for the parameters ($a_{1:p},\sigma^2$) are given by
\begin{align}
    a_{M} &= \left(\lambda\eye_p + \sum_{i=1}^{m}X^\top X\right)^{-1}\left(\lambda\mu_0 +  \sum_{i=1}^{m}X^\top x\right)\\
    \sigma^2_{M} &= \frac{2\beta + \lambda||a_{M}-\mu_0||^2 + \sum_{i=1}^{m}||x - Xa_{M}||^2}{2(\alpha+1) + mT + (m-1)p},
\end{align}
where recall our notation $X:=X_{p+1:T}$ and $x:=x_{p+1:T}$ and we have assumed that we have $m$ observations of the signal (though in practice usually only $m=1$ observation is available).

In this model, the hyperparameters can be chosen using a Grid search and cross-validation with a 5-fold split: for each set of hyperparameters, compute the MAP estimate on the train split, to then score it on the test split using the likelihood given that MAP for $a_{1:p}$ and $\sigma^2$ as the score function. The final score assigned to the hyperparameters is the average of the scores obtained on each fold. Lastly, the set of hyperparameters with the best score will be used in the posterior.

The grid to be used for the hyper-parameters is
\begin{align}
    \lambda, \alpha, \beta &\in \{0.1, 1, 10, 100\} \\
    \mu_0 &\in \left\{\begin{bmatrix}
           -10 \\
           -10 \\
           \vdots \\
           -10
         \end{bmatrix}, \begin{bmatrix}
           -8 \\
           -8 \\
           \vdots \\
           -8
         \end{bmatrix}, \hdots, \begin{bmatrix}
           10 \\
           10 \\
           \vdots \\
           10
         \end{bmatrix} \right\} \subset \mathbb{R}^p
\end{align}

% Add Case study part
%!TEX root = Bayesian_AR_LACCI2021.tex

\section{Simulations}
We validated the proposed Bayesian autoregressive spectral estimation method, BASE, over three case studies, two for synthetic data and one for real world data. The first one considered a synthetic AR signal, where the true and approximate model orders were different, this simulation is aimed to show that BASE is robust to model misspecification due to the chosen prior. The second experiment illuminates the unbiasedness and concentration properties of the approximate posterior over PSDs by applying BASE on a continuous-time signal generated by a Gaussian process (GP) \cite{Rasmussen:2006} with known PSD. Finally, the third experiment validates the ability of the proposed BASE model to detect periodicities from a sub-sampled real world astronomical time series. We also compare, over all the case studies, the performance of the method in both versions: the Model I in Sec.~\ref{sec:model_one} based on a half-Normal noise prior (using MCMC) and the Model II in Sec.~\ref{sec:model_one} based on the conjugate Inverse-Gamma noise prior with closed-form posterior. In the experiments using MCMC we used a NUTS sampler \cite{Homan:2014:NSA:2627435.2638586} in the PyMC3 Python toolbox \cite{pymc3}.

\subsection{Misspecified model order: AR-generated time series}
\label{sec:exp_AR}

\begin{figure}[H]
 \includegraphics[width=1\linewidth]{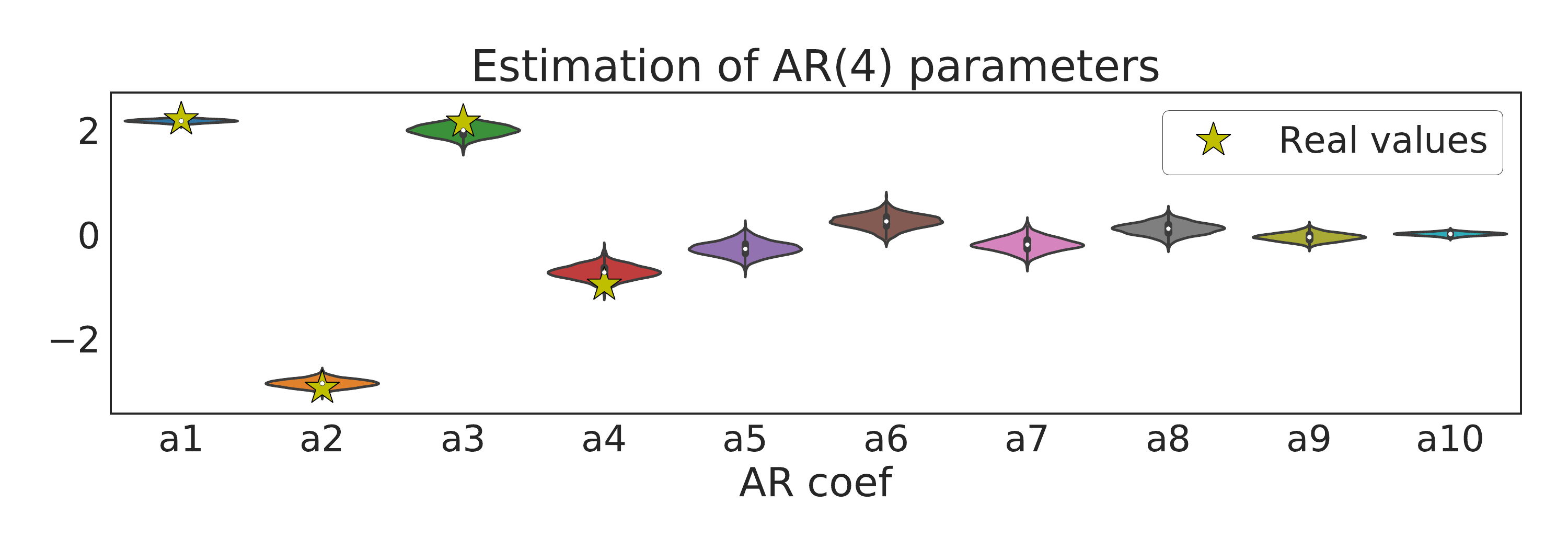}
 \caption{Approximate posterior distributions over AR parameters using MCMC (NUTS) and the true AR(4) values shown in yellow stars.}
 \label{im:ar_1}
\end{figure}

We implemented a stable AR(4) time-series to generate data that was later processed by BASE (Model I) with the assumption of a AR(10) model. Fig.~\ref{im:ar_1} shows the true values of the regression coefficients and their approximate posterior using MCMC; notice that these posteriors contain the true AR(4) coefficients, and they also converge to narrow posteriors centred around zero for higher orders coefficients. This validates the self-regularisation property, or robustness to misspecified model order, within the proposed BASE-framework which will be key when computing the sample PSDs.

\begin{figure}[t]
 \includegraphics[width=1\linewidth]{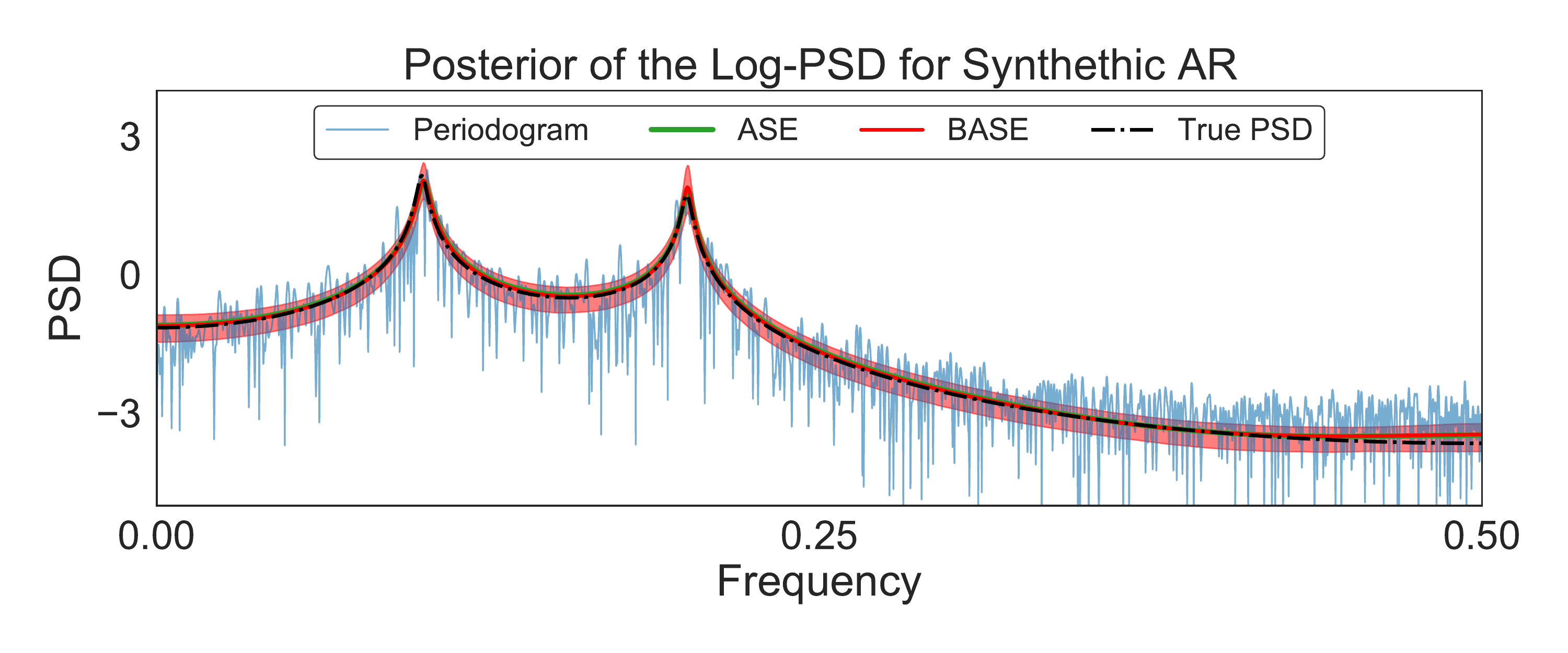}
 \caption{Log-PSD estimate for a synthetic AR(4) signal using the BASE method, true PSD (dashed black line), Periodogram (blue line), ASE (green line), and proposed BASE (red line, 95\% error bars in light red).}
 \label{im:ar_2}
\end{figure}

Next, using the approximate posterior over the AR parameters (coefficients and variance), we generated PSD samples according to eq.~\eqref{eq:arma_psd}. Fig.~\ref{im:ar_2} shows the density over PSDs using the proposed BASE (95\% error bars), the standard autoregressive spectral estimation (ASE), the true PSD and the Periodogram. Notice that the proposed BASE succeeded in estimating the true PSD while remaining unbiased and concentrated around the true value.  

\vspace{-.2cm}

\subsection{Continuous-time signal: Gaussian process with Laplace covariance}
\label{sec:exp_GP}

Let us consider a synthetic signal generated from a Gaussian process (GP) \cite{Rasmussen:2006} with Laplace covariance function given by 
\begin{equation}
 	K(\tau) = \sigma^2\exp(-|\tau|/l) \label{eq:laplace},
 \end{equation}
 where $\sigma^2$ denotes the marginal covariance and $l$ is known as the process \emph{lenghtscale}. Furthermore, we considered zero-mean Gaussian noise added to the GP sample. Recall that a sample from a GP can be regarded as a signal generated by an AR($\infty$) model. 

\begin{figure}[t]
 \includegraphics[width=1\linewidth]{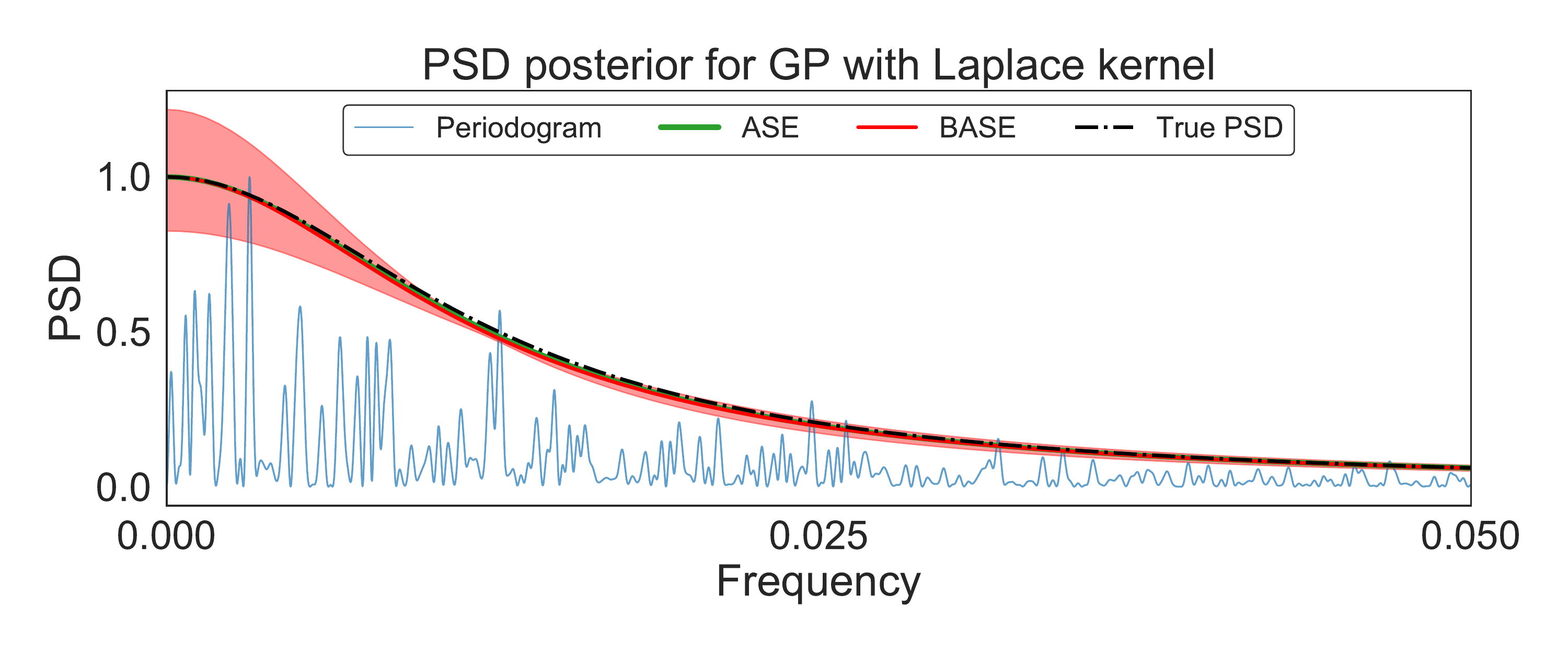}
 \caption{Spectral estimation of a GP trajectory: BASE shown in red, true PSD in dashed black, Periodogram in blue and ASE in green. The figure only shows the region [0, 0.05], since this is where almost all the spectral content of the data is contained. }
 \label{im:gp_1}
\end{figure}

 The proposed BASE (Model I) was implemented on this synthetic signal with an AR(4) model, the order of which was chosen from the rate of decay of the Laplace kernel. Then, the PSD estimate of the BASE model was compared against those of the standard ASE of the same order and the Periodogram. Fig.~\ref{im:gp_1} shows the PSD estimates (95$\%$ error bars for BASE) as well as the true PSD of the signal, given by the Fourier Transform of the Laplace kernel. Notice that the PSD posterior within BASE was concentrated around the true PSD and was able to capture the mean of its deterministic counterpart.

\subsection{Finding the periodicity of an astronomical time series}
\label{sec:exp_astro}

For the third experiment, we considered a real-world time series which may not be an AR process and therefore its PSD will not be precisely given by the expression in eq.~\eqref{eq:arma_psd}. This experiment was constructed to validate the Bayesian approach to handle model misspecification by averaging over the posterior models. The signal considered was the sunspots dataset \cite{sunspots}, known for having a period of 11 years (frequency $\approx0.0909[$years$^{-1}]$), we implemented BASE (Model I) to detect this periodicity only using $1/6$ of the available data. 

\begin{figure}[t]
 \includegraphics[width=1\linewidth]{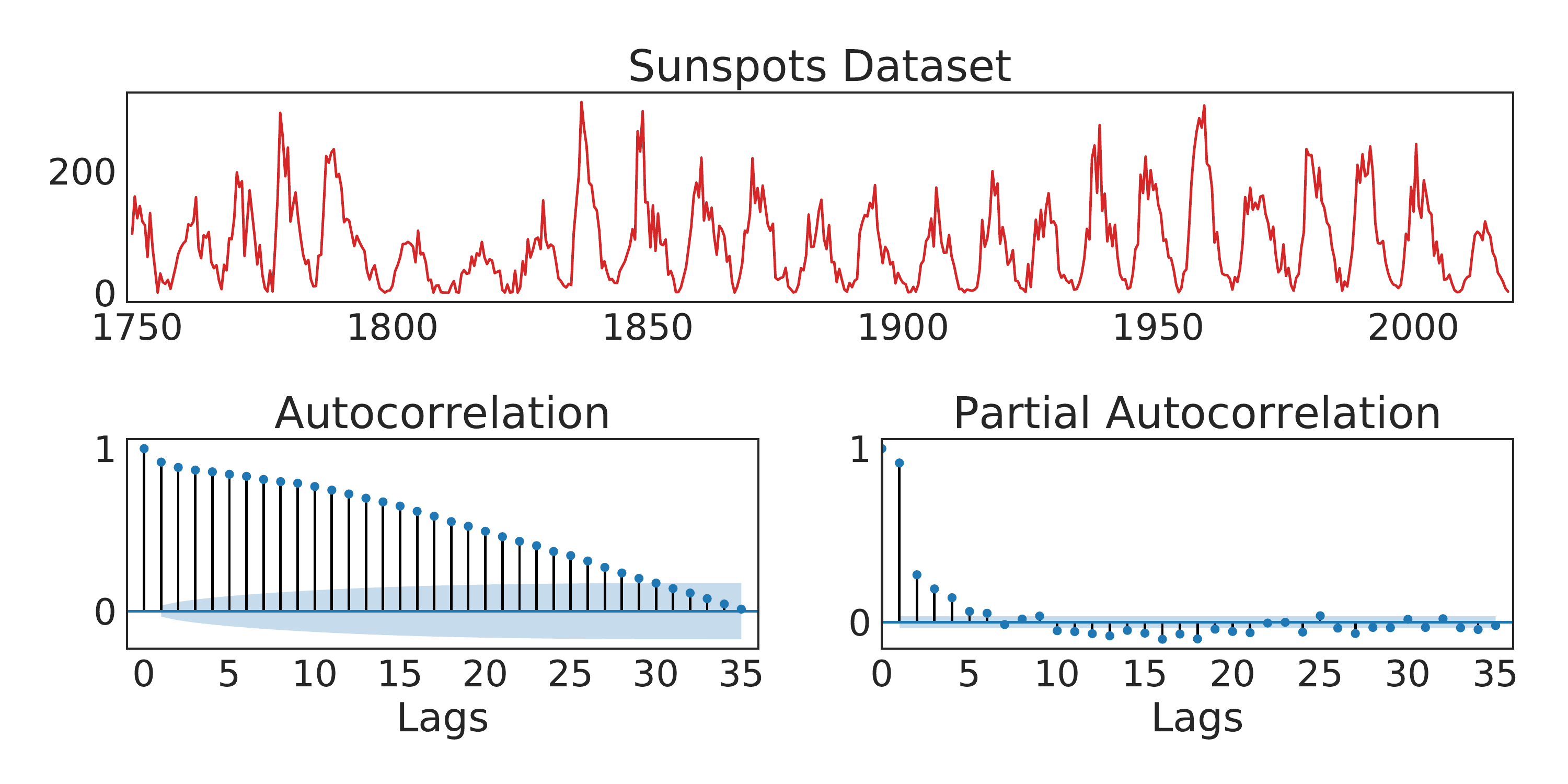}
 \caption{Sunspots time series (top) alongside autocorrelation and partial autocorrelation functions.}
 \label{im:sun_1}
\end{figure}

\begin{figure}[t]
 \includegraphics[width=1\linewidth]{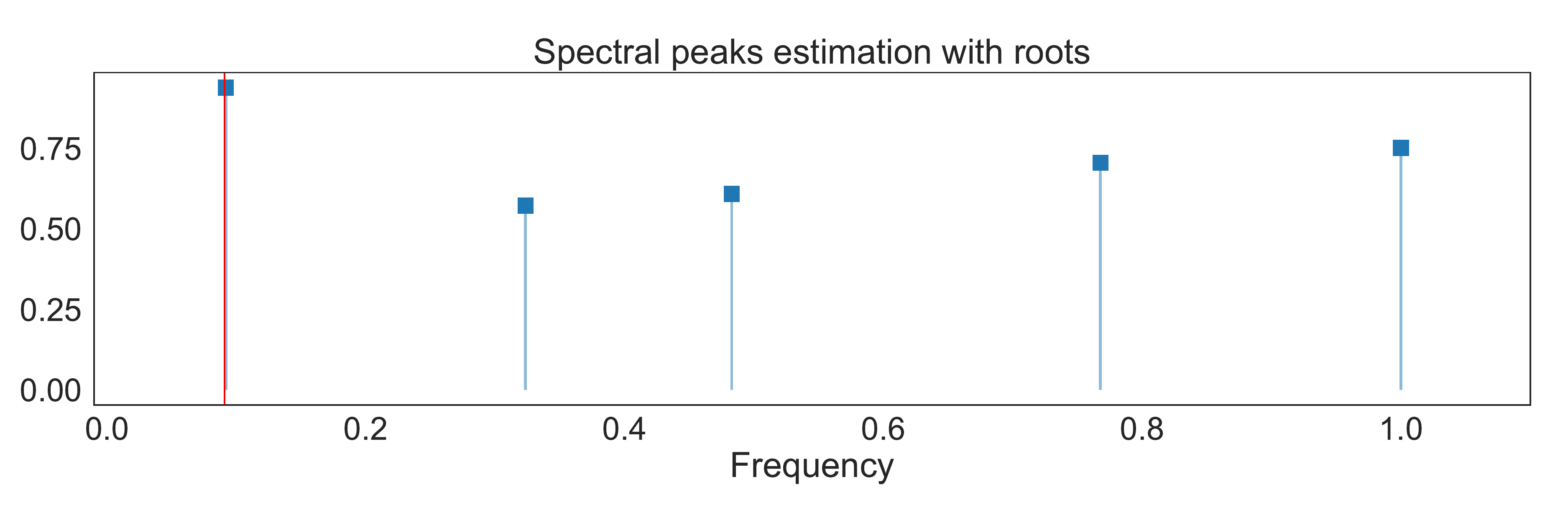}
 \caption{Peaks of the PSD computed by BASE for the sunspots time series. The red stem denotes the main peak.}
 \label{im:sun_peaks}
\end{figure}

\begin{figure}[t]
 \includegraphics[width=1\linewidth]{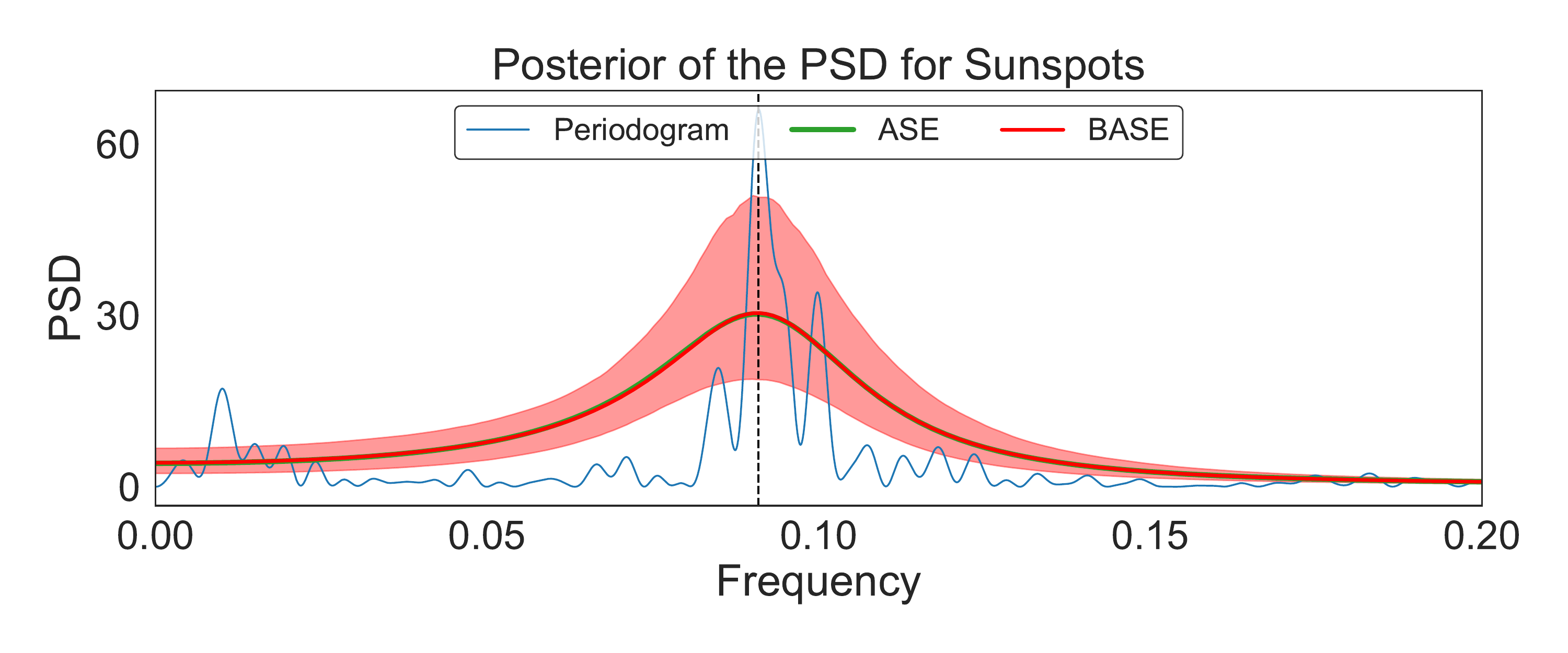}
 \caption{PSD estimates for the sunspots series: BASE (95\% error bars), ASE and Periodogram. The frequency of the well-known 11-year period is illustrated by a vertical dashed black line.}
 \label{im:sun_2}
\end{figure}

We first computed both the autocorrelation (ACF) and partial autocorrelation (PCF) functions of the sunspots series. Fig.~\ref{im:sun_1} shows the ACF and PCF, and show that due to the slow decay of PCF, the order of the autoregressive component of the signal cannot be specified with any certainty; this supports the evidence that the sunspot series is not an AR process. We implemented BASE (Model I) with an AR(9) model to then compare it against the ASE and the Periodogram, where ASE and the Periodogram used all the available data. Fig.~\ref{im:sun_peaks} shows the peak estimates of the PSDs using BASE, where the main peak (colour-coded in red) is $0.09181278[$years$^{-1}]$ and thus in line with the expected frequency peak. Fig.~\ref{im:sun_2} shows the PSD estimates for the three models,  with the proposed BASE model exhibiting its peak near the correct known period of the sunspots signal, as it was expected from Fig.~\ref{im:sun_peaks}. 

Notice that the errors bars of BASE are now larger than those of the previous experiments, this is because the signal considered in this experiment is not coming from an AR process and thus we are in the scenario of model mismatch. In this sense, due to the fact that BASE averages over possible AR models, its error bars are larger but they contain other peaks of the Periodogram. Finally, observe that despite the fact that an order $p=9$ for the model was chosen, this did not result in multiple peaks which validates BASE's robustness to overfitting arising from its Bayesian nature.

\subsection{Performance of the closed-form posterior}

BASE (Model II) was also implemented in the same experiments described above for comparison. Since this model has a closed form posterior, we refer to it as ``CF'' in the remaining part of this section. 

\begin{figure}[t]
 \includegraphics[width=1\linewidth]{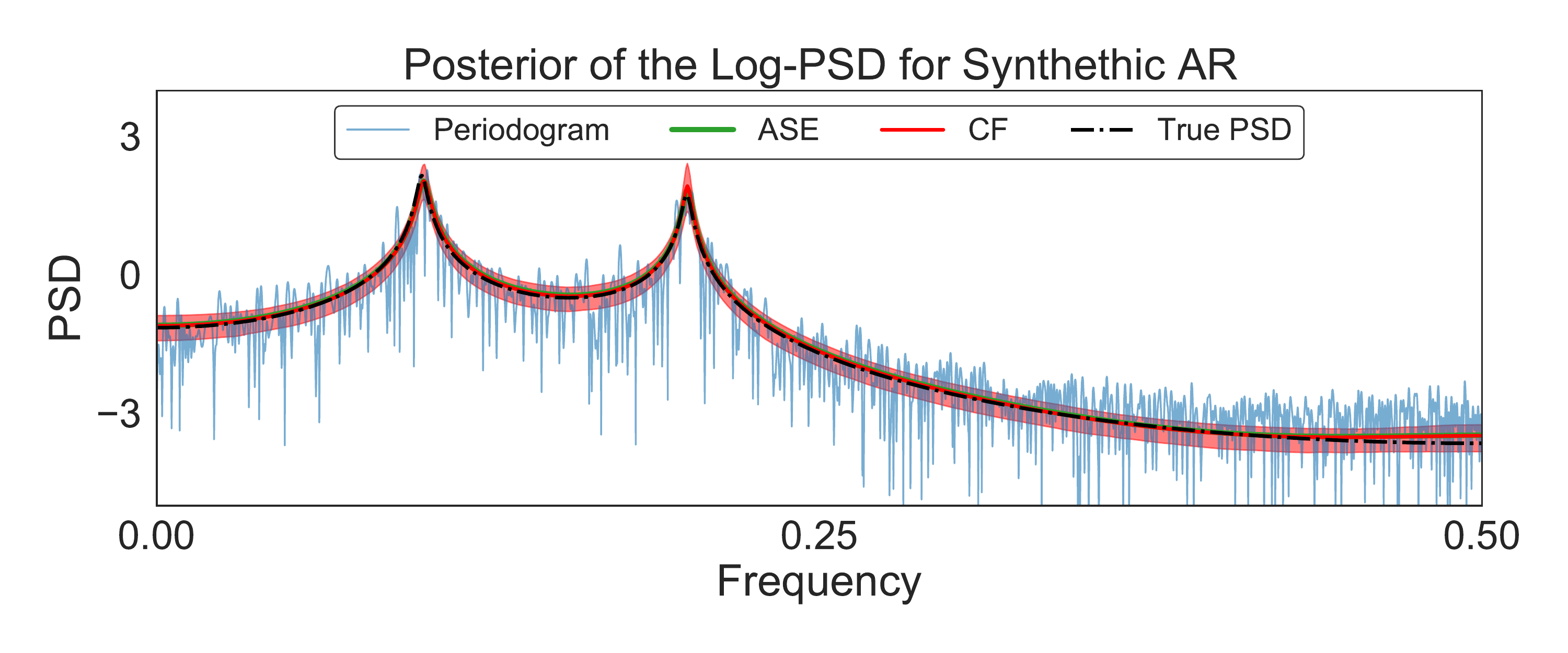}
 \caption{Log-PSD estimate for a synthetic AR(4) signal using the CF method; True PSD (dashed black line), Periodogram (blue line), ASE (green line), and proposed CF (red line, 95\% error bars).}
 \label{im:ar_3}
\end{figure}

Following the experiment in Sec.~\ref{sec:exp_AR}, the proposed CF method was implemented on the AR(4) synthetic time series, where the real parameters are known. Fig.~\ref{im:ar_3} shows the PSD estimate (95\% error bars) along with the ASE estimation, the true PSD and the Periodogram. By comparing these results with those shown in Fig.~\ref{im:ar_2}, note that the estimation of the mean is only slightly better, likewise, the variance of the estimation is slightly reduced.

\begin{figure}[t]
 \includegraphics[width=1\linewidth]{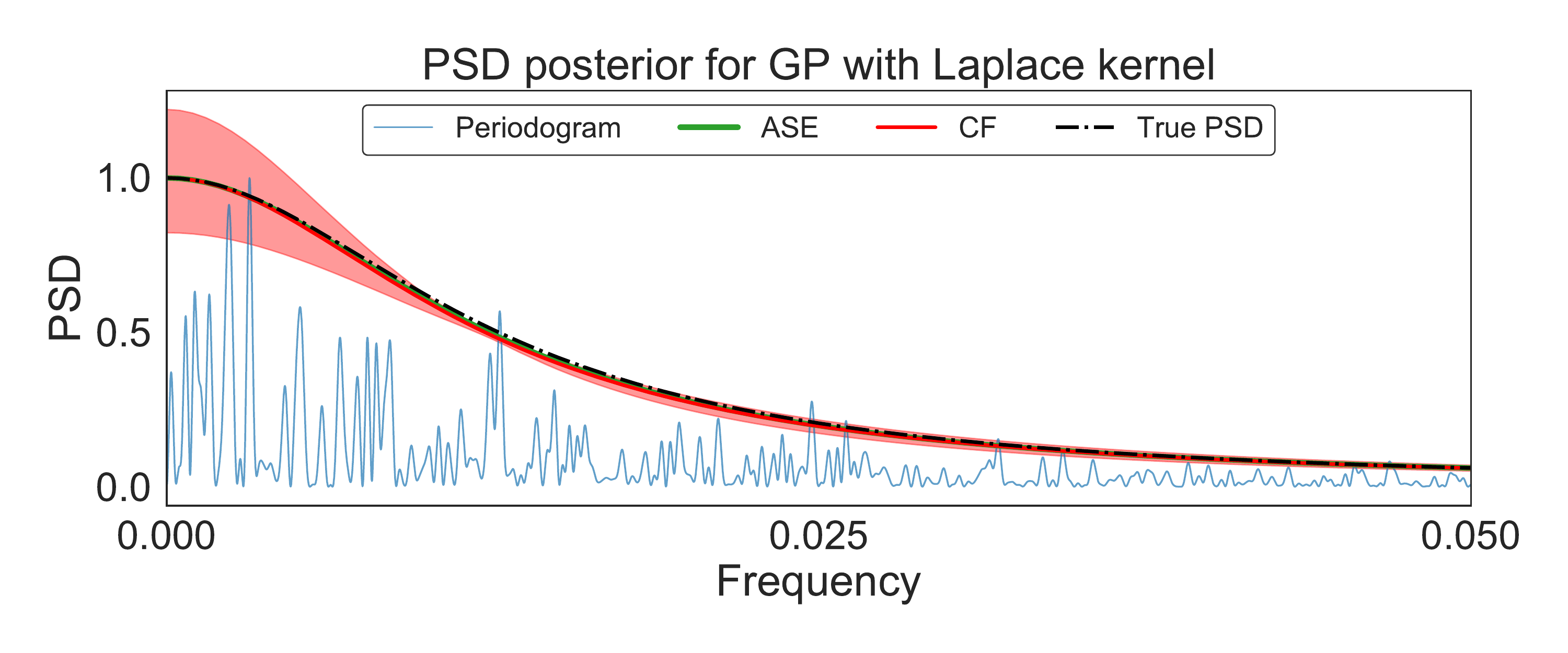}
 \caption{Spectral estimation of a GP trajectory: CF shown in red, true PSD in dashed black, Periodogram in blue and ASE in green. The figure only shows the region [0, 0.05], since this is where almost all the spectral content of the data is contained. }
 \label{im:gp_2}
\end{figure}

Next, we implemented CF on the same synthetic Gaussian process signal generated in \ref{sec:exp_GP}. Fig.~\ref{im:gp_2} shows the PSD estimates (95$\%$ error bars for CF) as well as the true PSD of the signal, given by the Fourier Transform of the Laplace kernel. The estimates provided by the CF method (BASE Model II) are very similar to those of BASE Model I, which uses MCMC, as they also concentrates around the true PSD and its mean captures the estimation of the ASE. Notice that the error bars are slightly better for CF than those given by BASE Model I. This suggest that the posterior distribution is more concentrated on the true model.  

\begin{figure}[t]
 \includegraphics[width=1\linewidth]{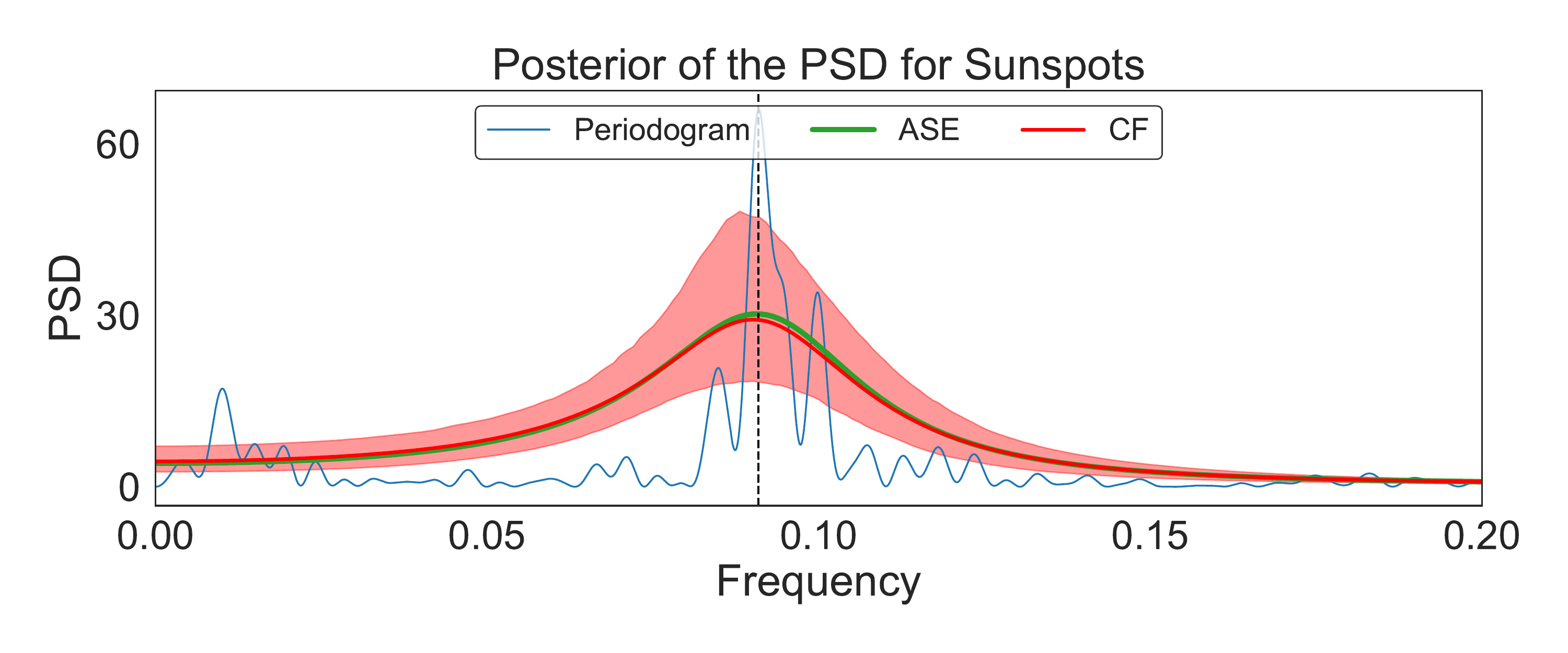}
 \caption{PSD estimates for the sunspots series: CF (95\% error bars), ASE and Periodogram. The frequency of the well-known 11-year period is illustrated by a vertical dashed black line.}
 \label{im:sun_3}
\end{figure}

Lastly, following Sec.~\ref{sec:exp_astro} we validated the CF model on real data by testing it on the sunspots data-set \cite{sunspots}. Fig.~\ref{im:sun_3} shows the results with the CF approach (BASE Model II), which again exhibits slightly less variance than BASE Model I using MCMC. Although the peak frequency found by CF has shifted to the left, the posterior PSD is still precise and concentrated on the true peak (found by computing the Periodogram over the entire dataset). This suggest that the CF could also recover the fundamental frequency of the data from a limited number of observations.

% Add discussion part
%!TEX root = Bayesian_AR_LACCI2021.tex

\section{Discussion and Future Work}

We have proposed a novel framework termed BASE, a Bayesian approach to autoregressive spectral estimation. BASE exploits the closed-form properties of the AR model to compute power spectral densities (PSD), and also complements it with a probabilistic treatment, thus allowing us to quantify uncertainty in the estimates of the PSDs through the use of confidence intervals. 

The description of the proposed method follows the idea that we can sample from the posterior distribution over PSDs (given a set of observations of the time series) by (i) sampling the posterior parameters of an AR model, to then (ii) compute the corresponding PSD of such sample parameter. In this setting, we proposed two models: both assume a multivariate normal prior on the AR parameters by they differ in that one assumes a half-Normal prior for the noise and the other one adopts an Inverse-Gamma; we dubbed them Model I and Model II respectively. BASE Model I uses MCMC to find the posterior samples while Model II features a conjugate prior and this results in a closed-form posterior. 

Both BASE models have been implemented on three experiments using: i) data generated by an AR model, ii) data generated by a Gaussian process with Laplace kernel, and iii) a real-world astronomical time-series. Through these simulations, we have validated BASE in terms of robustness to model misspecification, unbiasedness, accuracy, and periodicity detection. Regarding the comparison between Model I and Model II, we can say that, Model II gives posteriors that are tighter than those of Model I, arguably due to its use of a closed from posterior. Furthermore, the main feature of Model II is the reduction on the computation time due to avoiding the use of MCMC. The main setback of Model II, however, is that it requires the use of a specific prior.

Future work includes: i) heuristics to determine the hyper-parameters $\mu_0, \lambda, \alpha, \beta$, ii) a quantitative assessment of the reduction in computational complexity provided by Model II, iii) a development of a hierarchical prior to simultaneously identify the AR model order (e.g., a Dirichlet prior), and iv) extensions to ARMA models to cater for moving-average spectral components and spectra with zeros, iv) comparisons against nonparametric models such as \cite{tobar21b}.

\section*{Acknowledgements}
This work was funded by Fondecyt-Regular 1210606, ANID-AFB170001 (CMM) and ANID-FB0008 (AC3E).

% Below is an example of how to insert images. Delete the ``\vspace'' line,
% uncomment the preceding line ``\centerline...'' and replace ``imageX.ps''
% with a suitable PostScript file name.
% -------------------------------------------------------------------------
% \begin{figure}[htb]
%
%  \begin{minipage}[b]{1.0\linewidth}
%   \centering
%   \centerline{\includegraphics[width=8.5cm]{image1}}
%   %  \vspace{2.0cm}
%   \centerline{(a) Result 1}\medskip
%  \end{minipage}
%  %
%  \begin{minipage}[b]{.48\linewidth}
%   \centering
%   \centerline{\includegraphics[width=4.0cm]{image3}}
%   %  \vspace{1.5cm}
%   \centerline{(b) Results 3}\medskip
%  \end{minipage}
%  \hfill
%  \begin{minipage}[b]{0.48\linewidth}
%   \centering
%   \centerline{\includegraphics[width=4.0cm]{image4}}
%   %  \vspace{1.5cm}
%   \centerline{(c) Result 4}\medskip
%  \end{minipage}
%  %
%  \caption{Example of placing a figure with experimental results.}
%  \label{fig:res}
%  %
% \end{figure}

% To start a new column (but not a new page) and help balance the last-page
% column length use \vfill\pagebreak.
% -------------------------------------------------------------------------
%\vfill
%\pagebreak

% References should be produced using the bibtex program from suitable
% BiBTeX files (here: strings, refs, manuals). The IEEEbib.bst bibliography
% style file from IEEE produces unsorted bibliography list.
% -------------------------------------------------------------------------
\bibliographystyle{IEEEbib}
\bibliography{biblio}

\end{document}